# Eliminating Leakage in Volatile Memory with Anti-Ferroelectric Transistors

Hongtao Zhong[1,*], Zijie Zheng[2,*], Leming Jiao[2], Zuopu Zhou[2], Chen Sun[2], Xiaoyang Ma[3], Vijaykrishnan Narayanan[4], Huazhong Yang[1], Kai Ni[5,✉], Xiao Gong[2,✉] and Xueqing Li[1,✉]

[1]Tsinghua University, Beijing, China;

[2]National University of Singapore, Singapore;

[3]Princeton University, Princeton, NJ, USA;

[4]Pennsylvania State University, State College, PA, USA;

[5]Rochester Institute of Technology, Rochester, NY, USA;

[*]These authors contributed equally: Hongtao Zhong, Zijie Zheng;

[✉]To whom correspondence should be addressed; E-mail: kai.ni@rit.edu, elegong@nus.edu.sg, xueqingli@tsinghua.edu.cn

**Abstract**

Cache serves as a temporary data memory module in many general-purpose processors and domain-specific accelerators. Its density, power, speed, and reliability play a critical role in enhancing the overall system performance and quality of service. Conventional volatile memories, including static random-access memory (SRAM) and embedded dynamic random-access memory (eDRAM) in the complementary metal-oxide-semiconductor technology, have high performance and good reliability. However, the inherent leakage in both SRAM and eDRAM hinders further improvement towards smaller feature sizes and higher energy efficiency. Although the emerging nonvolatile memories can eliminate the leakage efficiently, the penalties of lower speed and degraded reliability are significant. This article reveals a new opportunity towards leakage-free volatile static memory beyond the known paradigms of existing volatile and nonvolatile memories. By engineering a double-well energy landscape with the assistance of a clamping voltage bias, leakage-free and refresh-free state retention of volatile memory is achieved for the first time. This new memory is highlighted by both the ultra-low leakage of nonvolatile memories and the speed, energy, and reliability advantages of volatile memories. A proof-of-concept memory is demonstrated using in-house anti-ferroelectric field-effect transistors (AFeFETs), delivering an extrapolated endurance of about $10^{12}$ cycles, a retention time of over 10 years, and no subthreshold channel leakage current. Such a new concept of AFeFET-based memory enables an improved balance between density, power, and reliability beyond all existing memory solutions.

**Introduction**

A large, fast, and reliable cache near the processing units boosts the system performance and energy efficiency for data-intensive computing. In the past few decades, the static random-access memory (SRAM) and the embedded dynamic random-access memory (eDRAM) have been the major adopted cache technologies to balance capacity, energy efficiency, reliability, etc. However, as the complementary metal-oxide-semiconductor (CMOS) technology scaling slows down, it has been increasingly challenging to provide larger cache capacity to meet the need of the unprecedented amount of data and computation demands by the emerging applications of Internet of Things[1], computer vision[2], cloud computing[3], etc.

Essentially, both SRAM and eDRAM in CMOS are volatile memories, and the leakage issue is a significant challenge that limits their performance and density scaling for future generations. For example, for SRAM, as the logic transistors approach the physical limits, the increasing subthreshold channel leakage current makes SRAM power-hungry[4-6]. Such an inherent physical mechanism hinders further improvement of the embedded cache capacity and also the performance in many low-power devices such as the Internet of Things and wearable gadgets[7-9]. It is also reported recently that the 3nm-node SRAM has very limited density improvement over the 5nm-node SRAM[10], indicating the challenge of the conventional CMOS process development for SRAM density scaling. For eDRAM, the continual device scaling towards higher density results in a severe challenge in maintaining the refresh rate. Meanwhile, although some emerging nonvolatile memories (NVMs) may potentially deliver higher density without leakage current issue, the inherent penalties of degraded speed and/or reliability are still limiting their application[11]. Given that, developing a new type of energy-efficient, fast, and reliable volatile memory is highly desired and is the focus of this work.

The excitation originates from the re-visiting of the fundamental memory state representation with the underlying energy landscape of a memory device. For both NVM and SRAM, as illustrated in Fig. 1**a**, the energy landscape can be approximated as a double-well potential, where the two local energy minima correspond to the two stable states. Such a landscape is shaped by the powered-on inverter loop in an SRAM at the cost of leakage currents or the inherent energy barrier between states in NVM devices. Initialized at one state, the memory switches to the other state upon a write pulse as the energy landscape tilts and the initial state is no longer stable. After the removal of the write pulse, the double-well energy landscape is restored and the memory stays at the new state.

Differently, in the case of DRAM, including the embedded DRAM (eDRAM), as illustrated in Fig. 1**b**, the most stable state is when the capacitor is free of charge, which corresponds to the energy landscape with a single energy minimum. When applied with a write pulse to charge up the capacitor, a temporary double-well potential emerges and switching to the other state proceeds. However, with the removal of the external write pulse and the dominance of the leakage current, such a transient energy landscape collapses back into a single energy minimum upon removal of the write pulse. Consequently, refresh operations become necessary to preserve the DRAM memory states.

A fundamental question has thus emerged: is it possible to engineer the volatile memory devices to have a double-well energy landscape, and maintain the memory states without the costs of refresh activities and static holding power? Inspired by this question, this work proposes a new class of leakage-free volatile memory (LFVM), featured with ultra-low-power refresh-free static data retaining capabilities. It exploits the double-well energy landscape with the assistance of a clamping voltage. With the assisting clamping voltage applied to the device gate, the stable double-well energy landscape is maintained. This concept is shown in Fig. 1**c**, in which a write-access transistor ($T_w$) and a hysteretic device are illustrated. The gate of the hysteretic device and the bitline connected to $T_w$ are both biased at $V_m$. Thanks to the stable hysteresis of the hysteretic device, the two memory states can be maintained. One highlight of this new memory concept is that, with the same drain and source voltage in $T_w$, there is no subthreshold leakage current between the drain and the source during the idle state. This feature leads to ultra-high standby energy efficiency. Another highlight is the simplified peripheral control and circuitry. This is because no refresh operation is needed, and the memory-state-independent assisting gate bias can be applied globally to all cells in the array. Lastly, this general memory structure has a much higher density than SRAM with fewer transistors, while also providing the static random access capability.

Given those intriguing highlights, this article further explores the design of candidate hysteretic devices for the proposed LFVM concept. It is exciting to see that both anti-ferroelectric field-effect transistors (AFeFETs) and nanoelectromechanical (NEM) relays fit well with the goal. NEM relays rely on the electrostatically actuated mechanical switching mechanisms[12,13]. The combination of the elastic force, the van der Waals force, and the electrostatic force exhibits the desired hysteresis in Fig. 1c. Currently, NEM relays are being actively developed to integrate and scale with the CMOS technology. In this article, the proof of concept is based on AFeFETs[14,15]. As a derivative of ferroelectric FET (FeFET), AFeFET integrates an anti-ferroelectric (AFE) film into the gate stack of a metal-oxide-semiconductor field effect transistor (MOSFET), which exhibits the hysteresis in the positive gate voltage region, as shown in Fig. 1c, with a high ON/OFF ratio, low write power, and compatibility with advanced CMOS (theoretical physical mechanisms are included in the supplementary materials). More importantly, the AFE devices exhibit a significantly improved endurance over their ferroelectric (FE) counterparts, which has been experimentally demonstrated by several recent works[15-17]. Actually, the main challenge for AFeFET in replacing FeFET is the volatility of AFE materials, where the remnant polarization is zero upon the removal of the external gate bias. Although techniques such as DRAM-like periodic refresh or built-in electric field strategy[17,18] can be adopted, the overheads are high. The former approach comes with pricy refresh power consumption and normal access stalls. The latter of engineering built-in bias into the AFE thin film faces constraints by limited tuning range and flexibility[17,18]. This is because the work functions of available CMOS-compatible metals are difficult to cover a wide range, especially considering the Fermi-level pinning. Therefore, the proposed LFVM solves the volatility issue of existing AFeFET memories, and provides a new opportunity to exploit the benefits of high endurance, scalability, and energy efficiency.

Leveraging the appropriate AFE and channel materials, the proposed AFeFET can be integrated within the back-end-of-line (BEOL) thermal budget (≤450 °C), thus allowing multi-tier memory stacking. In this work, we experimentally demonstrate a BEOL-compatible AFeFET using a low thermal budget, wide bandgap, and high electron mobility amorphous-indium-gallium-zinc-oxide (a-IGZO) channel and the BEOL-compatible and scalable $Hf_{1-x}Zr_xO_2$ FE/AFE material system. Our AFeFETs were realized using the metal-ferroelectric-metal-insulator-semiconductor (MFMIS) structure with the flexibility to optimize the gate stack electric field, featuring an extrapolated endurance of about $10^{12}$ cycles, a retention time of more than 10 years with the proposed LFVM operation, a high ON/OFF ratio, and small device-to-device variation. In contrast to the previous impression of impracticality in using AFE devices for memory purposes, we believe that the proposed LFVM opens up a new paradigm for power-efficient and high-performance AFE memory.

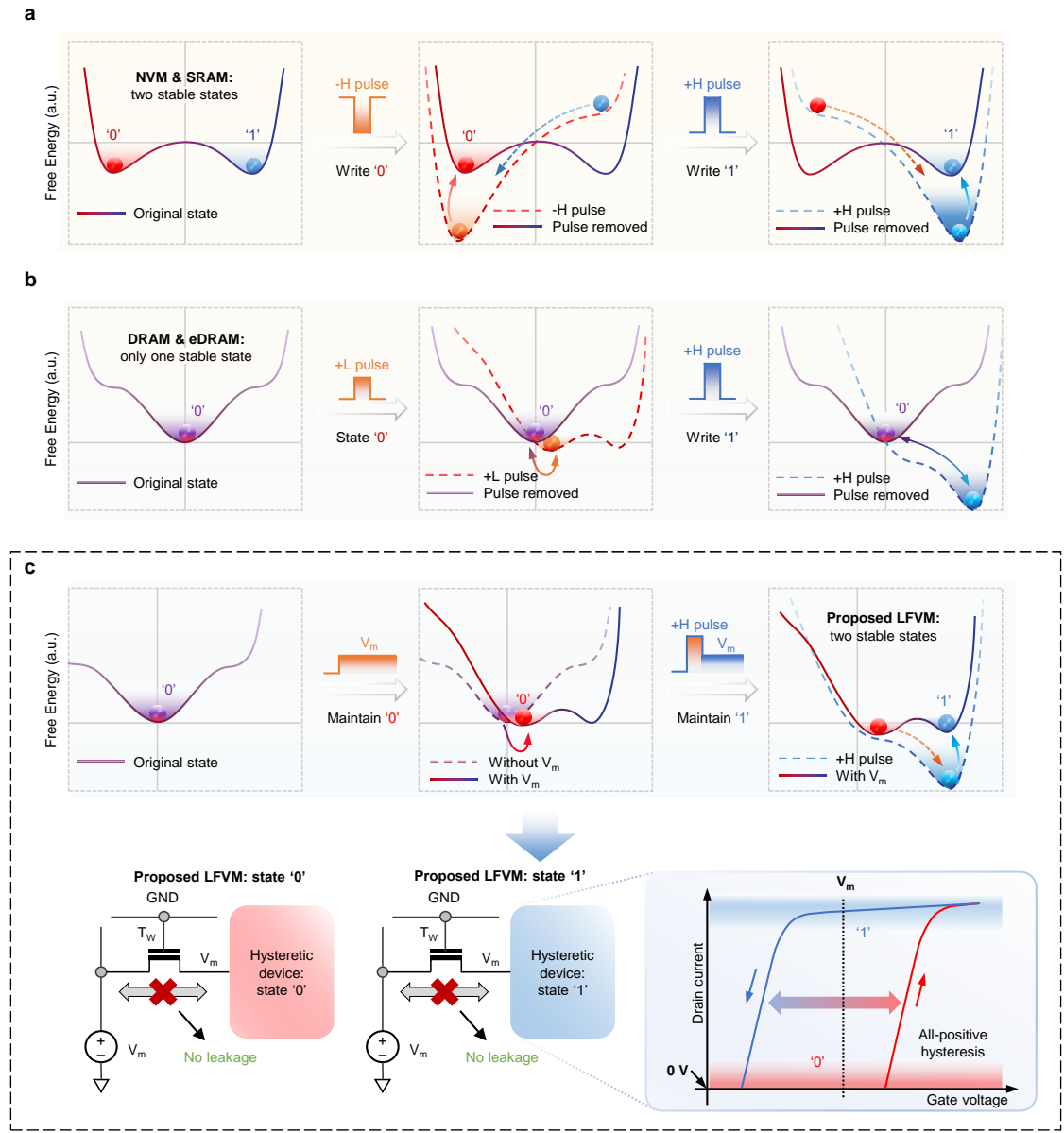

**Fig. 1 | Physical mechanism of different memories**. **a**, State maintaining and switching methods of NVM and SRAM: For NVM, the energy landscape is double-well with two local minimum energy states '0' and '1'. To switch between the states, a high negative or positive pulse is applied. After the pulse is removed, the state will be stabilized to the local minimum energy state '0' or '1'. The energy landscape of SRAM is also double-well but the state switching for SRAM involves VDD and GND only. **b**, State maintaining and switching methods of DRAM and eDRAM: There is only one local minimum energy state '0'. To switch from '0' to '1', a high positive pulse is applied. Returning to '0' involves GND only and does not need a high negative pulse. It is observed that the required voltage range is relatively small. However, the state '1' is unstable and it will return to '0' eventually due to leakage, if the external pulse

is removed. Therefore, periodic refresh is adopted in practice to maintain the data. **c**, Proposed LFVM memory based on the hysteretic volatile devices. For a category of volatile devices, with the proposed LFVM operation, both states '1' and '0' can be maintained using the same voltage configuration. Besides, within one such memory cell, it consumes no subthreshold drain-source leakage power, thanks to the same applied voltage at the drain and the source of the turned-off write access transistor. This is achieved by exploiting the volatile hysteresis memory window.

**LFVM Operation with AFeFET in 2T1AF Memory Cell**

To reveal how AFeFETs can be exploited to enable LFVM, the AFeFET operation principles are discussed first. It leverages the versatile $Hf_{1-x}Zr_xO_2$ material system, where the parameter $x$ yields mixed FE and AFE characteristics[19]. When $x$ increases from 0.5 to 1, $Hf_{1-x}Zr_xO_2$ with strong ferroelectricity will become more dominated by anti-ferroelectricity with Zr-rich composition. As shown in Fig. 2**a**, the polarization-voltage (*P-V*) curves measured on fabricated AFE capacitors show a typical volatile behavior with pinched hysteresis. For Zr-rich $Hf_{1-x}Zr_xO_2$, such characteristic originated from a phase transition induced by the external electric field, which can be observed in both positive and negative branches[20,21]. As introduced earlier, DRAM-like periodic refresh and built-in bias introduction have been reported to address the volatility issue of capacitor-based memory. In addition to the challenges introduced earlier, the effectiveness of those approaches remains unclear for transistor-based memory. By replacing the normal gate dielectric with an AFE thin film, typical AFeFETs can be formed, which also exhibit volatile characteristics. Like the measured drain current to gate voltage ($I_D$-$V_{GS}$) characteristics for a fabricated IGZO AFeFET shown in Fig. 2**b**, an anti-clockwise hysteresis is observed during the 0 V – 4 V – 0 V sweep. When $V_{GS}$ returns to 0 V, the high current state is gone, and the next sweep follows the same path with a more positive threshold voltage ($V_{TH}$), indicating volatility. In practice, the hysteresis window might be limited in the 0 V – 4 V – 0 V sweep, and a small negative voltage, i.e., -2 V in this article, can be applied to enlarge the hysteresis window for a larger initial ON/OFF ratio. The main reason is due to the non-zero voltage drop in the AFE layer with $V_{GS}$ = 0 V, which will be further analyzed in the next section. To address the volatility issue, in this work, we will apply the proposed LFVM operation and demonstrate its intriguing memory operation.

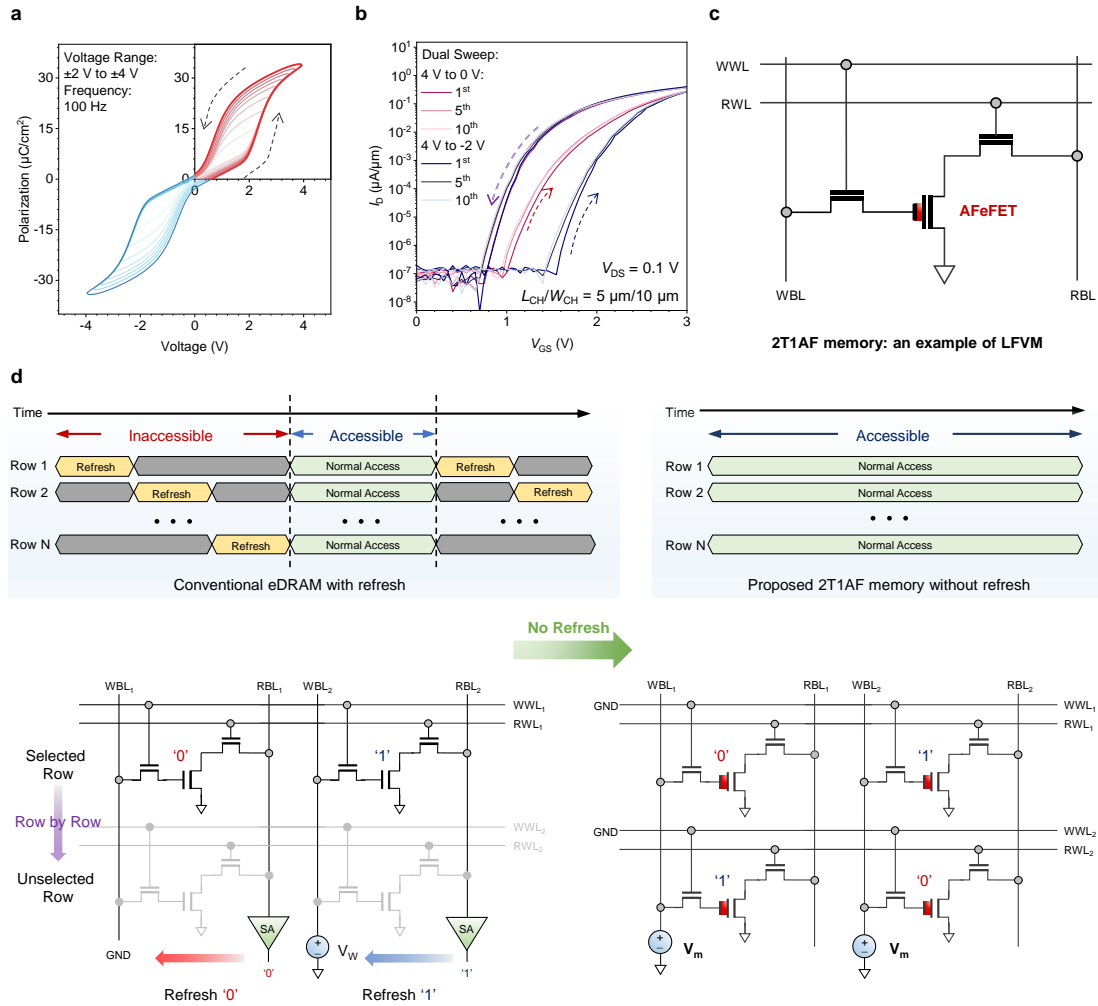

**Fig. 2 | Device electric characteristics and 2T1AF memory. a**, P-V curves of the Hf$_x$Zr$_{1-x}$O$_2$-based AFE capacitor. There is a double-hysteresis loop, and the polarization returns to zero at zero voltage, which shows the volatility. **b**, $I_D$-$V_{GS}$ curves measured on the fabricated a-IGZO AFeFETs. The hysteresis loops for the 1$^{st}$, 5$^{th}$, and 10$^{th}$ cycle of measurement are shown with repeated and all-positive hysteresis windows, which also implies volatility. **c**, Proposed 2T1AF memory. The write access transistor is controlled by the write word line (WWL), and the input write datum is sent through the write bit line (WBL) to program the AFeFETs. The 2T1AF memory cell evolves from the conventional 3T eDRAM cell with the transistor for data storage replaced by an AFeFET. The 2T1AF memory benefits from non-destructive read and negligible read and write disturbance with the extra access transistors. **d**, Difference in data maintaining method between the 3T eDRAM and the 2T1AF memory. The 3T eDRAM utilizes the conventional refresh operation to maintain data. The refresh operation reads out data and then writes them back row by row. The refresh operation not only consumes high refresh power, but also stalls the normal access requests. The 2T1AF

memory utilizes the LFVM operation to maintain data. The states '0' and '1' share the same voltage configuration, so all data stored in the memory array can be maintained simultaneously. The LFVM operation not only consumes ultra-low power but also stalls no access requests. Therefore, the accessible time window for the memory array is extremely extended.

From the array perspective, the LFVM operation in AFeFETs can be understood and evaluated. While denser memory cells are possible, a 2-transistor-1-AFeFET (2T1AF) disturb-free memory cell is taken as an example to illustrate the LFVM operation, as shown in Fig. 2**c**. The 2T1AF cell consists of a write access transistor, a data storage transistor, and a read access transistor. Its structure is similar to that of 3T eDRAM, with the transistor for data storage replaced by an AFeFET. The write and read operations of the 2T1AF memory cell are also similar to those of the 3T eDRAM cell. Detailed voltage configuration is shown in Supplementary Fig. 2 and Supplementary Fig. 3. The use of write access transistor and read access transistor allows non-destructive read, and prevents write and read disturbance. The peripherals can detect the voltage change on the read bitline (RBL) to sense the stored data.

The main difference between the 3T eDRAM and the 2T1AF memory is the data retention mechanism. The 3T eDRAM utilizes the periodic two-step refresh operation consisting of data read and write-back, as illustrated in Fig. 2**d**. Because the stored data are usually different between rows, the refresh operation is executed row-by-row. Conventional refresh operation not only consumes high refresh power, but also occupies the access time window and degrades the memory performance due to the access stalls. As the memory size increases, the refresh power consumption and performance degradation due to refresh become worse[22]. On the contrary, the 2T1AF memory utilizes the proposed LFVM operation to maintain data. Unlike the refresh operation, the LFVM operation can maintain the cell states '0' and '1' using the same bias voltage configuration. Therefore, the read-and-then-write-back refresh operation is not necessary, and all data stored in the memory array can be maintained. Besides, the LFVM operation itself consumes no subthreshold drain-source leakage and does not stall normal access. This feature saves power and prevents refresh-originated performance degradation, which is highly desired by many low-power applications.

# AFeFET and 2T1AF Memory Demonstration

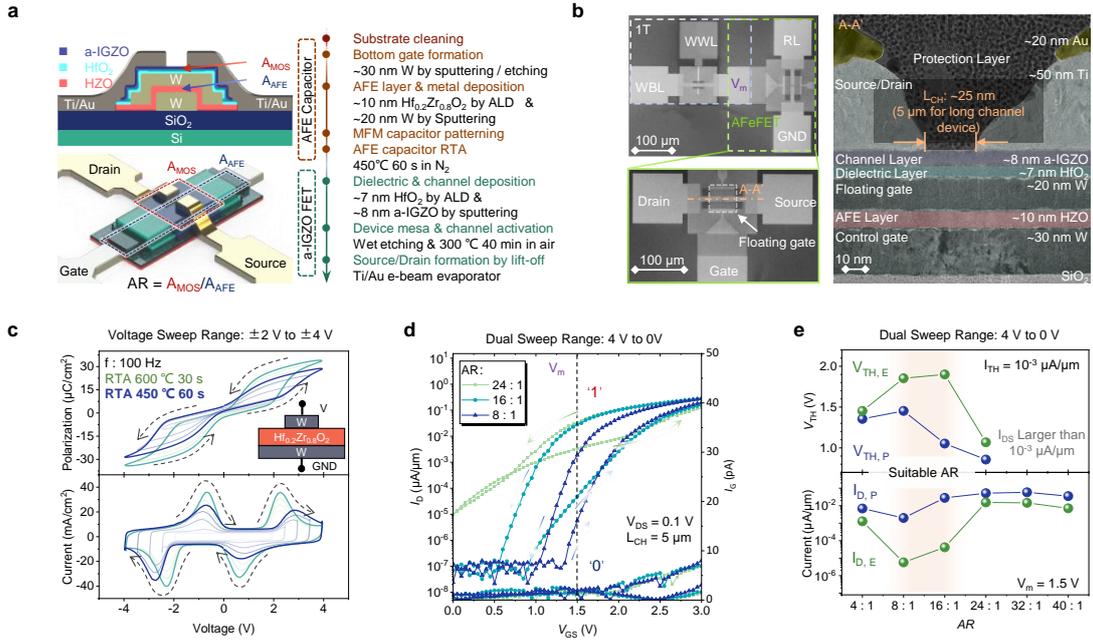

**Fig. 3 | Fabricated device and memory cell with basic characterizations. a**, Device structure and key process steps of the a-IGZO AFeFETs. The cross-sectional view illustrates a gate stack of IGZO/HfO$_2$/W/HZO/W from top to bottom. The schematic illustration utilizes the MFMIS structure. $A_{MOS}$ and $A_{AFE}$ are areas of the AFE layer and the MOSFET, respectively, and the voltage distribution across the gate stack can be tuned by changing the area ratio (AR) of $A_{MOS}$ to $A_{AFE}$. $L_{CH}$ represents the gap between the drain and the source. **b**, The top-view SEM image of the a-IGZO AFeFET and 2T1AF memory, and the TEM image in the channel region with the thickness of each layer. The read access transistor is regarded as a part of the source measure unit (SMU) to detect the AFeFET current. **c**, The *P-V* double-hysteresis loops and the *I-V* loops of the AFE capacitor with different RTA temperatures. Typical AFE characteristics are validated for both conditions. **d**, $I_D$-$V_{GS}$ curves of the a-IGZO AFeFETs with 5 μm $L_{CH}$ and various ARs. The anti-clockwise hysteresis can be observed with the volatile property induced by the AFE layer within the gate stack. **e,** The $V_{TH}$ with fixed current criteria and $I_D$ at $V_m$ of the a-IGZO AFeFETs with various ARs. The characteristics of the AFeFETs can be tuned by AR, and 8:1 and 16:1 are suitable for the proposed memory cell with a relatively large memory window (MW) and a large ON/OFF ratio.

Fig. 3**a** summarizes the cross-sectional view, the schematic illustration and the key process steps of the fabricated a-IGZO AFeFET (more details are referred to Supplementary Fig. 4). The AFeFETs

use the MFMIS structure with the flexibility to engineer the area ratio (AR) of the MOSFET area ($A_{MOS}$) to the AFE area ($A_{AFE}$) to tune the overall characteristics of the device. Besides, the staggered bottom-gate structure with a top-contacted drain/source is utilized, and $L_{CH}$ is determined by the gap between the drain and the source. The a-IGZO thickness is about 8 nm, and the gate stack comprises 7 nm $HfO_2$ (dielectric layer)/20 nm W (floating gate)/10 nm $Hf_{0.2}Zr_{0.8}O_2$ (AFE layer)/30 nm W (control gate) from top to bottom. Fig. 3**b** shows the scanning electron microscope (SEM) image of the fabricated AFeFET device and the 2T1AF memory cell. The read access transistor is regarded as a part of the source measure unit (SMU) to detect the AFeFET current, so only the write access transistor and the AFeFET device are fabricated to demonstrate the functionality of the LFVM operation. The thicknesses of each layer mentioned above in the channel region (A-A') are confirmed by the high-resolution transmission electron microscopy (HRTEM) image.

Fig. 3**c** illustrates the *P-V* and *I-V* characteristics of the AFE capacitors with a MIM structure to show the anti-ferroelectricity. The capacitors are fabricated using the same process as the AFeFETs, as shown in Fig. 3**a**. Note that, in the *P-V* curves, in contrast to the FE capacitors, there is negligible remnant polarization ($P_r$) at 0 V, implying the volatility of AFE materials. In the *I-V* curves, a double current peak in both positive and negative branches corresponding to a pinched hysteresis can be observed. Meanwhile, the rapid thermal annealing (RTA) with both 450 ℃ 60 s and 600 ℃ 30 s can provide strong AFE behavior, indicating the BEOL compatibility of the HZO for AFE applications. With relatively higher annealing temperature, the voltage for AFE to FE phase transition becomes smaller, which might lead to lower operation voltage, as shown in Fig. 3**c**. Fig. 3**d** shows the $I_D$-$V_{GS}$ curves of the a-IGZO AFeFETs ($L_{CH}$ = 5 μm, $V_{DS}$ = 0.1 V) with various ARs of 8:1, 16:1, and 24:1 under dual sweep measurement from 4 V to 0 V. Due to the electrically coupled AFE capacitor on the MOSFET gate, the $I_D$-$V_{GS}$ curves of the AFeFETs exhibit anti-clockwise hysteresis with volatile properties.

Tuning AR is an effective approach to optimize the device characteristics for the proposed memory design, which is equivalent to adjusting the capacitance ratio between the AFE MIM capacitor and the MOS capacitor. Because the two capacitors are in series, the electric field distribution in the AFE thin film can be modulated. As a result, a larger AR results in a larger voltage drop across the AFE layer, leading to a wider hysteresis window or memory window (MW) in the $I_D$-$V_{GS}$ curves, where MW is defined as the $V_{TH}$ difference between $V_{TH, P}$ of the programmed state

'1' and $V_{TH, E}$ of the erased state '0', at a constant current of $10^{-3}$ μA/μm.

However, although a sufficient AR is necessary for a large MW, the experimental results indicate that the AR of suitable AFeFETs should be selected within a reasonable range for the proposed LFVM. As shown in Fig. 3**d**, the devices with a large AR of 24:1 exhibit a low ON/OFF ratio that is defined as the ratio of $I_D$ at the programmed state to the erased state. This can be understood with consideration of the capacitance matching between the MIM AFE capacitor and the MOS capacitor. The small absolute polarization of AFE capacitors with a small area cannot create enough change in the channel carrier density. As a result, there exists an optimal AR range to maximize both the MW and the ON/OFF ratio simultaneously. More detailed $I_D$-$V_{GS}$ and $C$-$V_{GS}$ characteristics of the fabricated AFeFETs are referred to Supplementary Fig. 5.

Fig. 3**e** summarizes $V_{TH}$ and $I_D$ at $V_m$ of 1.5 V for both states with various ARs. 1.5 V is set for $V_m$ with consideration of the impact of $V_m$ on both the ON/OFF ratio and the leakage power referred to Supplementary Fig. 6. Here, the data are extracted from the abovementioned $I_D$-$V_{GS}$ curves to show the impact of AR on MW and the ON/OFF ratio. It can be observed that as AR increases within a certain range, MW and the ON/OFF ratio improve because of more saturated switching of the AFE layer. The AFeFETs with AR from 8:1 and 16:1 satisfy the proposed memory cell design, where MW and the ON/OFF ratio of ~1 V and ~700 for the fabricated devices, respectively, can be realized.

Moreover, due to the non-zero flat-band voltage in the gate stack and inevitable charge trapping within the floating gate, it is possible that the back switching from the FE phase to the AFE phase can be displaced from $V_{GS}$ = 0 V. In some cases, a negative gate bias may be needed to reverse the FE phase back to the AFE phase. Therefore, as shown in Supplementary Fig. 5**a**, MW without negative biases is relatively narrower in most cases, corresponding to the minor loop operation of the AFeFETs, as illustrated in Supplementary Fig. 7**c**. Hence, a negative write voltage for erase operation will be helpful for the memory operation to enable a large ON/OFF ratio and a large MW.

Moreover, the device variation and scalability are further explored. We found that the fabricated a-IGZO AFeFETs exhibit excellent device-to-device uniformity, and scaled devices maintain proper function with $L_{CH}$ down to ~50 nm with boosted $I_D$ and negatively shifted $V_{TH}$. The experimental results are referred to Supplementary Fig. 8.

## Measurement and benchmarking of the 2T1AF Memory

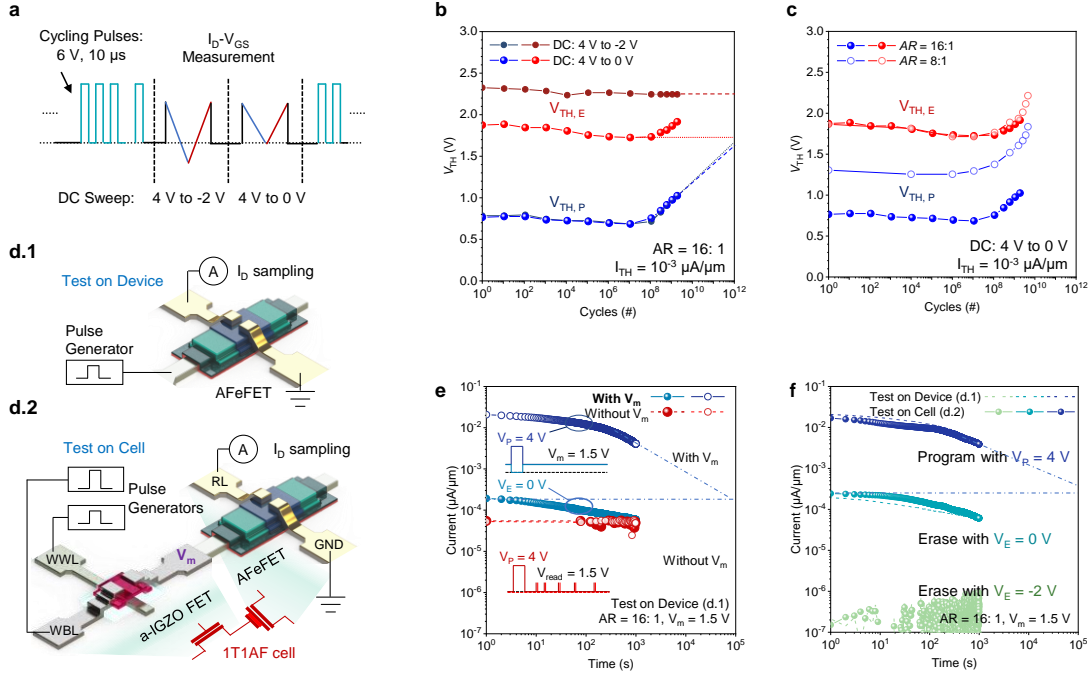

**Fig. 4 | Endurance and retention measurement of the AFeFETs and the memory cell. a**, Schematic of the endurance measurement using unipolar cycling pulses. The DC sweep ranges include both 4 V to -2 V and 4 V to 0 V. **b**, Endurance measurement of the AFeFETs with AR of 16:1, showing high endurance of about $10^{12}$ cycles with the 4 V to -2 V operation and $10^{10}$ cycles with the 4 V to 0 V operation. Here, the endurance limit is defined as the extrapolated MW degradation to half the original. **c**, Endurance measurement of the AFeFETs with ARs of 8:1 and 16:1. The endurance property of the devices with 16:1 AR is better due to the relatively wider MW. **d**, Schematic of the retention measurement for testing on both a single AFeFET device and the LFVM cell. **e**, Retention measurement of the AFeFETs with and without $V_m$ during the standby period. Without $V_m$, the AFeFETs cannot be used as a static memory due to the volatile and dynamic characteristics of the AFE layer, as shown with the two red lines. Here, the read voltage is set to 1.5 V, equal to $V_m$. In contrast, the blue lines are obtained by applying $V_m$ = 1.5 V and sampling $I_D$ after the program and the erase operations. Two distinct states can be observed with the proposed operation scheme. **f**, Retention measurement in the memory cell showing results similar to the device retention and proving a successful operation of the proposed memory cell. With the unipolar operation, the retention time is estimated to be longer than $10^4$ s. By applying a low negative voltage ($V_E$ = -2 V) for erase operation, the extrapolated retention time over 10 years can be achieved.
12

The combination of AFE materials and the proposed LFVM operation yields a memory device with good reliability. Fig. 4a illustrates the endurance cycling waveform applied to the gate of the AFeFETs. Unipolar cycling pulses of 6 V and 10 μs are applied to the device gate, followed by the DC sweep ranging from both 4 V to -2 V and from 4 V to 0 V, to obtain the $I_D$-$V_{GS}$ curves. Fig. 4b shows the measurement results of the AFeFETs with an AR of 16:1. Our devices maintain proper function after more than $10^9$ cycling without dielectric breakdown or significant MW degradation. This proves that using the AFE material with unipolar stressing for the devices can significantly improve the endurance compared with its FE counterpart[23]. Generally, charge injection into the FE or AFE layer is regarded as one of the primary culprits causing fatigue, and several studies attribute the higher endurance of AFE material to the lower local depolarization field during the switching transient that prevents severe injection effects[24,25].

Besides, as mentioned above, a larger MW can be observed with a DC sweep from 4 V to -2 V because of the more saturated switching of the AFE layer. If we quantify the measured endurance limit when the extrapolated MW is degraded by half, the endurances for the 4 V to -2 V operation and the 4 V to 0 V operation are in the order of $10^{12}$ and $10^{10}$, respectively. Fig. 4c shows the endurance measurement of the AFeFETs with AR of both 16:1 and 8:1. With a larger MW, the endurance of the devices with AR of 16:1 is improved. It also shows a parallel $V_{TH}$ shift for both states, suggesting that the absolute $V_{TH}$ shift is the main degradation, rather than the MW shrinking. Since there is no significant imprint of the AFE capacitor as shown in Supplementary Fig. 9, it is more likely caused by the charge injection to the floating electrode or bias temperature instability (BTI) of the IGZO channel, which causes parallel $V_{TH}$ shift for both states. These might be mitigated with more careful device design and optimized device fabrication.

Fig. 4d depicts the schematic of the retention measurement setup at both the device level and the memory cell level. To ensure the proper write and read operation for both the device and the memory cell, the voltage pulses utilized here have a fixed pulse width of 10 ms, which is long enough to charge the capacitor and switch the AFE layer. The program pulse amplitude is 4 V, while the erase pulse amplitude varies from 0 V to -2 V. For the proposed design, after each memory write, a hold bias $V_m$ = 1.5 V is applied to hold the memory state. This bias amplitude is selected to avoid significant AFE switching during the hold period so that proper operation with a relatively large ON/OFF ratio and long retention time can be realized. Fig. 4e depicts the retention measurement results with and without $V_m$ during the standby period. It is evident that without $V_m$, the AFeFETs lose the programmed state, and cannot be used as a memory, as suggested by the merged red curves. However, by applying $V_m$, the memory states can be kept in AFeFETs, and two distinct states can be observed with a decent ON/OFF ratio and long retention time.

At last, we demonstrate a successful operation of the memory cell with similar voltage conditions while the write access transistor is turned on during the write operations and turned off during the idle hold time. The transistor gate is set to be floating to eliminate the gate leakage (~10 pA) from the write access transistor during the retention measurement. Fig. 4**f** shows that with the unipolar operation, the retention time is less than $10^5$ s, shorter than the case where $V_E$ = -2 V is applied for memory erase. As mentioned above, it is mainly due to the non-saturated depolarization of the AFE layer at 0 V gate voltage. With the applied negative erase pulse, $I_D$ in the erase state ($I_{D, E}$) decreases and the ON/OFF ratio increases. Therefore, the memory retention improves accordingly with over 10 years of extrapolated retention time. The experimental results have verified the feasibility of the proposed LFVM operation and device design based on AFeFETs. Further scaling and process optimization can be conducted to achieve better performance in terms of speed and energy efficiency.

One potential application of the 2T1AF LFVM is the embedded memory integrated with the logic circuits on the same chip for data-intensive edge-computing applications[7-9]. Unlike the stand-alone memories, embedded memories can provide lower access latency but are more limited by the die area. The most common embedded SRAMs (eSRAMs) feature fast speed, mature fabrication, good reliability, high standby leakage power and a large cell footprint of 120-150 $F^2$, where F is the feature size. The eDRAMs, including the capacitor structures (e.g., 1T1C[26] and 2T1C[27]) and transistor-only structures (e.g., 2T[28], 3T[29,30] and 4T[31]), are denser but face the challenges of high power dissipation and performance degradation due to frequent refresh operations.

The embedded NVM (eNVM) can eliminate leakage efficiently. There have been several NVM technologies being produced or developed, including the floating-gate (FG) transistors[32], the spin transfer torque magnetic RAMs (STT-MRAM)[33], the phase change RAM (PCM)[34,35], the resistive RAM (RRAM)[36,37], the ferroelectric capacitors (FeRAM)[38,39], and FeFET[40-42], etc. These NVM cells can store at least two states even when power is off by harnessing different physical mechanisms. However, these devices also face challenges. The FG transistors face the challenges of limited scalability, high voltages, and low endurance. The resistive NVMs, including STT-MRAM, PCM and RRAM, have high write power and limited write performance because of the current-driven write mechanism. Besides, large variations and a low ON/OFF ratio of the resistive devices make it difficult to distinguish different states, further increasing the error rate[43,44]. FeRAM and FeFET use an electric field and consume no DC power during the write operation. Therefore, the write energy can be significantly reduced. However, FeRAM suffers from the destructive read operation, while FeFET suffers from relatively low endurance[45] and switching stochasticity with device scaling[46].

As summarized in Table 1, the demonstrated 2T1AF memory shows a good balance among various metrics, including density, disturbance, power consumption and endurance. The 2T1AF memory has only 3 transistors without additional capacitors, so the area of the 2T1AF memory is much smaller than the conventional 6T SRAM. With the write and read access transistors, the 2T1AF memory also features non-destructive read operation. Besides, thanks to the DC-free capacitive load, the AFeFET write energy is also ultra-low. With the LFVM operation, no periodic refresh or normal access stall exists in the 2T1AF memory, and the actual standby power approaches zero unless with a non-ideal leaky gate. More detailed comparisons with SRAM and eDRAM are included in Supplementary Fig. 10. Furthermore, the measured endurance of the in-house AFeFET can be more than $10^9$ cycles, while the extrapolated endurance can reach $10^{12}$ cycles, much larger than optimized FeFETs with the same channel structure ($\sim 10^8$ cycles)[23].

It is important to point out that the proposed 2T1AF memory based on AFeFET is just one viable implementation of the LFVM memory. Other designs with different hysteretic devices, e.g., NEM relays[12,13], or different memory circuit topologies, e.g., 2T structure without the read access transistor, can also be explored in the future. Although the fabricated AFeFET device in this article was targeted at demonstrating functionality, rather than optimal performance, recent AFE-related works have shown that the AFE-based devices can embrace high speed, high endurance, low operating voltage, long retention time, etc[15,17]. Besides, the AFE-based memory can also be utilized to achieve multi-bit storage for even higher density[47]. Furthermore, the LFVM memory can also be applied to applications including compute-in-memory (CiM)[48], DNA sequence alignment[49], etc.

| | eSRAM | eDRAM (capacitor)[26,27] | eDRAM (transistor)[28-31] | FeRAM[38,39] | AFeRAM[15-17] | FeFET[40-42] | Proposed 2T1AF memory |
|---|---|---|---|---|---|---|---|
| Cell structure | 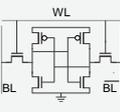 | 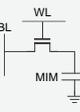 | 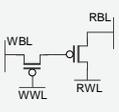 | 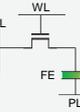 | 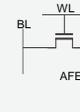 | 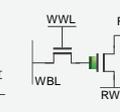 | 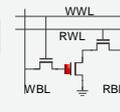 |
| Status | Available [a] | Available | Available | Available | Research [b] | Research | **Research** |
| Cell size | 120-150 $F^2$ | 40 $F^2$ | 40-60 $F^2$ | 30-40 $F^2$ | 30-40 $F^2$ | 10-30 $F^2$ | **60 $F^2$** |
| Non-volatile | No | No | No | Yes | Yes/No (with/without built-in field) | Yes | **No** |
| Non-destructive read | Yes | No | Yes | No | No | Yes | **Yes** |
| Write voltage | <1 V | <1 V | <1 V | 1-3 V | 1-3 V | 1.5-4 V | **4 V** |
| Write energy | ~1 fJ | ~1 pJ | ~1 fJ | ~100 fJ | ~100 fJ | 1-10 fJ | **1-10 fJ** |
| Standby power | High | Medium | Medium | Low | Low | Low | **Low** |
| Write speed | <1 ns | >10 ns | 1-5 ns | 1-10 ns | 1-10 ns | 1-10 ns | **1-10 ns** |
| Read speed | <1 ns | >10 ns | 1-5 ns | 1-25 ns | 1-25 ns | 1-10 ns | **1-10 ns** |
| Retention | Static [c] | Refresh | Refresh | 10 years | 10 years/Refresh (with/without built-in field) | 10 years | **10 years (static with ideal LFVM)** |
| Endurance | >$10^{16}$ | >$10^{16}$ | >$10^{16}$ | >$10^{12}$ | $10^9$-$10^{12}$ | $10^5$-$10^9$ [d] | **~$10^{12}$** |

[a] Available: The memory has been produced or manufactured.

[b] Research: The memory is still being researched.

[c] Static: The stored data keep maintained with a power supply.

[d] Some FeFETs with special structures (e.g., the back-gated structure[50]) can achieve >$10^{10}$ cycles.

Table 1. Performance comparison between different embedded memories.

## Conclusion

In this article, we propose a novel leakage-free volatile memory (LFVM) based on hysteretic devices with all-positive hysteresis. From the energy landscape perspective, the proposed LFVM memory has a stable double-well energy landscape with a holding voltage bias, which can eliminate the excess subthreshold drain-source leakage current and retain the advantages of the volatile memories. At the cell level, we also propose corresponding LFVM operation that can maintain stored data without the need of refresh operations. We have experimentally demonstrated the LFVM memory in a 2T1AF cell topology based on BEOL-compatible a-IGZO AFeFETs. The proposed 2T1AF memory has much higher endurance than FeFETs by harnessing the inherent AFeFET device capability. Besides, the 2T1AF memory also stands out with ultra-long retention time. Therefore, the LFVM memory opens a new category for a series of volatile memories, and is promising for low-power and endurance-sensitive scenarios.

## Acknowledgment

**Author contributions**

X. Li, X. Gong and K. Ni proposed and supervised the project. X. Li and H. Zhong conceived the LVFM design with corresponding operations. H. Zhong conducted the circuit-level simulations. Z. Zheng performed the device fabrication and the experimental verification. All authors contributed to the discussion and data analysis. H. Zhong, Z. Zheng, X. Li, X. Gong and K. Ni wrote the manuscript.

**Competing interests**

The authors declare no competing interests.

**Data availability**

The data that support the figures within this paper and other findings of this research are available from the corresponding authors with reasonable requests.

**Supplementary Materials**

**Theoretical analysis based on the LGD theory for FeFETs and AFeFETs**

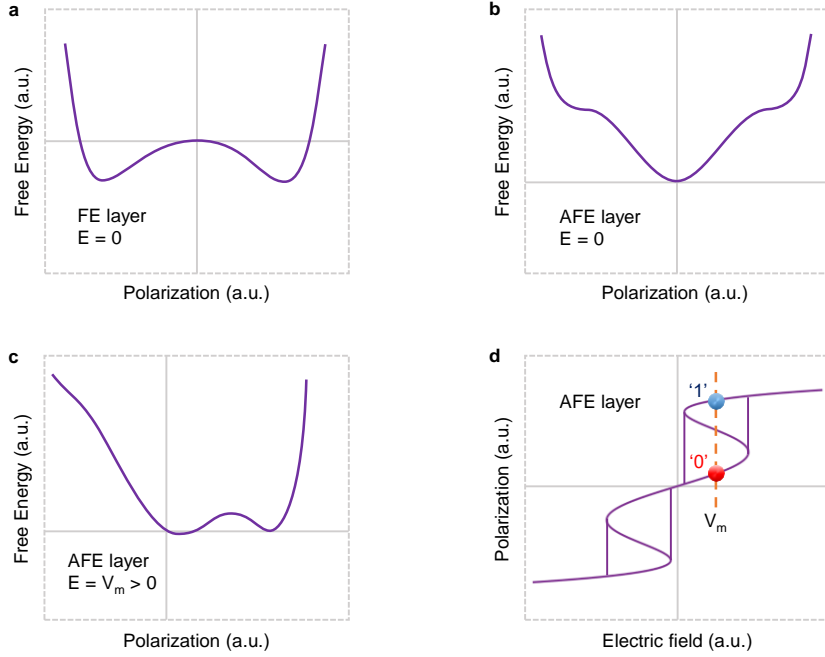

**Supplementary Fig. 1. a**, The free energy vs. polarization landscape of the FE layer with $E = 0$. **b**, The free energy vs. polarization landscape of the AFE layer with $E = 0$. **c**, The free energy vs. polarization landscape of the AFE layer with $E = V_m > 0$. **d**, The polarization vs. electric field landscape of the AFE layer.

Physical mechanisms of the FE layers and AFE layers in this article are based on the Landau-Ginzburg-Devonshire (LGD) theory, and the relationship between the Gibb's free energy ($G$) and the polarization ($P$) is as follows:

$$G = \left(\frac{\alpha}{2}\right)P^2 + \left(\frac{\beta}{4}\right)P^4 + \left(\frac{\xi}{6}\right)P^6 - E \quad (1)$$

where $E$ is the total electric field including the internal electric field and external electric field, and $\alpha$, $\beta$, and $\xi$ are constants. In the FE-based devices, $\alpha < 0$ for the FE layer. In the AFE-based devices, $\alpha > 0$ for the AFE layer. Supplementary Fig. 1**a** shows the free energy vs. polarization landscape of the FE layer with $E = 0$. There are two local minimum energy states without a power supply, which means the FE layer is nonvolatile. Supplementary Fig. 1**b** shows the free energy vs. polarization landscape of the AFE layer with $E = 0$. Different from the double well potential of the FE material, there is only one local minimum energy state at $P = 0$, which implies that the AFE layer is volatile. Supplementary Fig. 1**c** shows the free energy vs. polarization landscape of the AFE layer with $E = V_m > 0$. With the applied external $E$, there are two local minimum energy states. However, these two energy states are both unstable and will return to the local minimum energy in Supplementary Fig. 1**b** if the external field is removed.

Supplementary Fig. 1**d** shows the polarization vs. electric field landscape of the AFE layer. The landscape is obtained by minimizing the Gibb's free energy with a given $E$, i.e.,

$$\frac{\partial G}{\partial P} = \alpha P + \beta P^3 + \xi P^5 - E = 0 \tag{2}$$

$$\therefore P + \beta P^3 + \xi P^5 = E \tag{3}$$

Therefore, when $E = V_m > 0$ is within the hysteresis window, the solution of equation (3) includes two stable solutions and one unstable solution. The unstable solution is discarded and the remaining two solutions correspond to the two local minimum energy states in Supplementary Fig. 1**c**.

**2T1AF memory write operation**

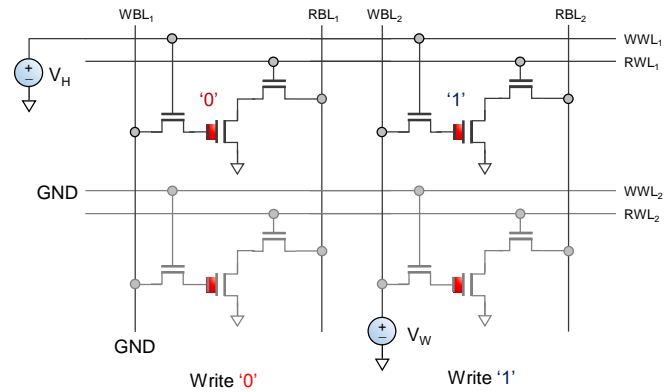

**Supplementary Fig. 2.** Write operation of the 2T1AF memory: writing '0' and '1'.

The write operation of the 2T1AF memory is similar to that of the conventional 3T eDRAM. The main difference is that the operating voltage of the 2T1AF memory is relatively large. As shown in Supplementary Fig. 2, to carry out the write access, the WWL in the selected row is firstly switched to a high write voltage ($V_H$) that can ensure the saturated polarization switching for the AFeFETs (i.e., $V_H > V_W + V_{TH}$). Other WWLs in unselected rows are grounded. Then, the WBLs are driven to $V_W$ (GND) if the data to be written to the 2T1AF memory cell are logic '1' ('0'). Note that the charges on the capacitance between the gate and the source ($C_{GS}$) of the AFeFETs represent the stored states. The $V_W$ in this article is 4 V.

**2T1AF memory read operation**

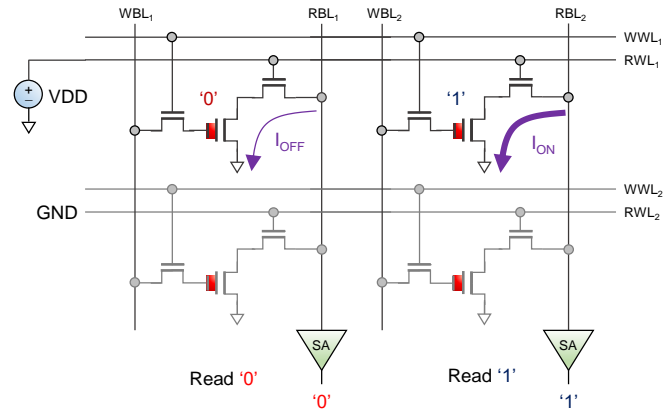

**Supplementary Fig. 3.** Read operation of the 2T1AF memory: reading '0' and '1'.

The read operation of the 2T1AF memory is also similar to that of the conventional 3T eDRAM. Unlike the write operation, the high voltage is not required in the read operation, and a low-voltage VDD is enough. As shown in Supplementary Fig. 3, to carry out the read operation, the RBLs in all columns are firstly pre-charged to VDD, and then the RWL in the selected row is driven to VDD to turn on all read access transistors in this row. Other RWLs in the unselected row are grounded. If the stored bit in a cell is '0', which means that the AFeFET of this cell is turned off, the pre-charged RWL will stay at VDD. In this way, the sense amplifier (SA) detects the OFF-state current ($I_{OFF}$) and outputs the result '0'. Otherwise, if the stored bit in a cell is '1', which means that the AFeFET of this cell is turned on, the pre-charged RWL will be pulled down to GND, and the SA detects the ON-state current ($I_{ON}$) and outputs the result '1'.

**Device fabrication**

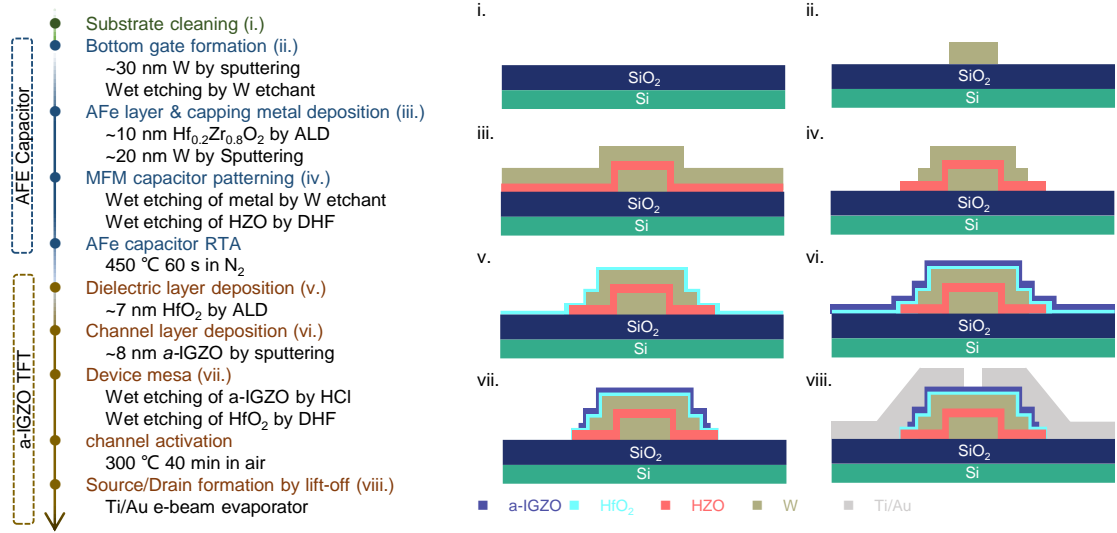

- Substrate cleaning (i.)
- Bottom gate formation (ii.)
  ~30 nm W by sputtering
  Wet etching by W etchant
- AFe layer & capping metal deposition (iii.)
  ~10 nm $Hf_{0.2}Zr_{0.8}O_2$ by ALD
  ~20 nm W by Sputtering
- MFM capacitor patterning (iv.)
  Wet etching of metal by W etchant
  Wet etching of HZO by DHF
- AFe capacitor RTA
  450 ℃ 60 s in $N_2$
- Dielectric layer deposition (v.)
  ~7 nm $HfO_2$ by ALD
- Channel layer deposition (vi.)
  ~8 nm a-IGZO by sputtering
- Device mesa (vii.)
  Wet etching of a-IGZO by HCl
  Wet etching of $HfO_2$ by DHF
- channel activation
  300 ℃ 40 min in air
- Source/Drain formation by lift-off (viii.)
  Ti/Au e-beam evaporator

**Supplementary Fig. 4.** Process steps of the a-IGZO AFeFETs. The memory cell can be formed with the access thin film transistor fabricated during the process flow of a-IGZO FET shown above. Devices with proper AR can achieve a non-hysteretic $I_D$-$V_{GS}$ characteristic and serve as the access transistor.

## $I_D$-$V_{GS}$ and $C$-$V_{GS}$ measurement

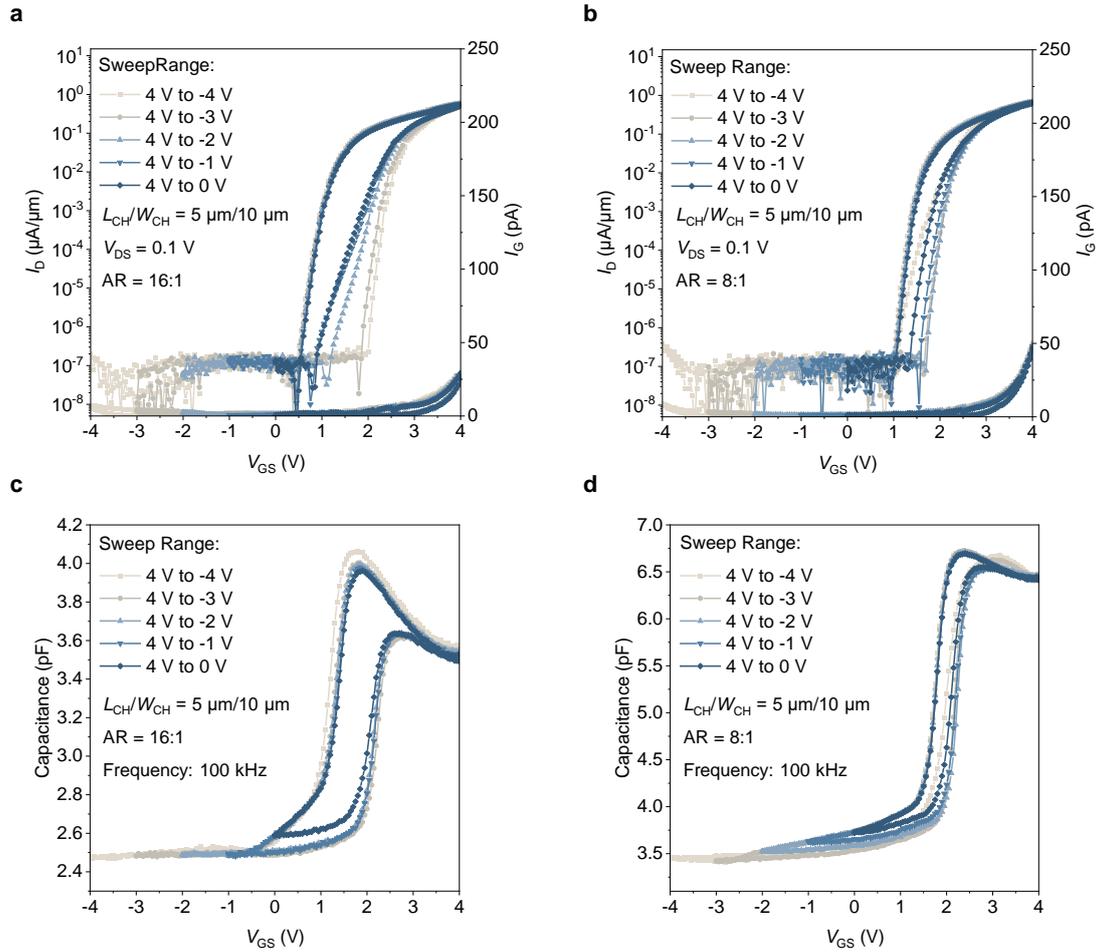

**Supplementary Fig. 5 | $I_D$-$V_{GS}$ and $C$-$V_{GS}$ of the a-IGZO AFeFETs. a** and **b** are the $I_D$-$V_{GS}$ curves of the AFeFETs with AR of 16:1 and 8:1, respectively. The gate bias voltage is from 4 V to 0 V and to different negative biases. **c** and **d** are the $C$-$V_{GS}$ curves of the AFeFETs with AR of 16:1 and 8:1, respectively. The capacitance measurement is conducted with one probe placed at the gate referring to the gate bias and two probes placed at the source and drain connected to the virtual ground. The small signal has an amplitude of 30 mV, while the gate bias voltage is from 4 V to 0 V or to different negative voltage biases. For the AFeFETs with AR of 16:1, the MW decreases with less negative voltage, which can be attributed to the partial switching of the AFE layers. For the devices with AR of 8:1, similar phenomena can be observed except for the case where -4 V is applied. Such decreased MW with a sweeping range from 4 V to -4 V can be induced by the tunneling between the channel to the floating gate.

**$V_m$ impact**

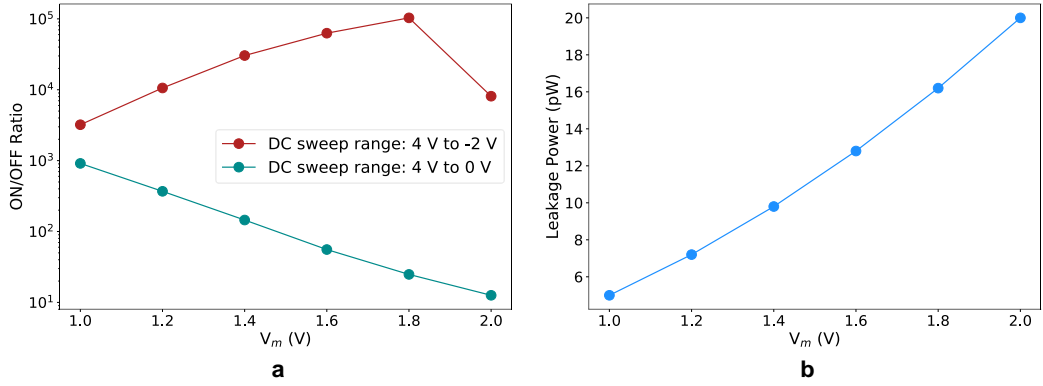

**Supplementary Fig. 6. a**, $V_m$ impact on the ON/OFF ratio. **b**, $V_m$ impact on the leakage power.

In the LFVM operation, the only requirement is that $V_m$ should be within the hysteresis window with some margin. Considering the noise and variation issues, the available $V_m$ range is smaller than the hysteresis window. Under this constraint, $V_m$ can be tuned and optimized to meet different design demands. In this section, we mainly discuss the $V_m$ impact on the ON/OFF ratio and the leakage power (i.e., the standby power). Supplementary Fig. 4**a** shows the $V_m$ impact on the ON/OFF ratio using the DC sweep ranging both from 4 V to -2 V and from 4 V to 0 V. As illustrated in Fig. 3, the DC sweep from 4 V to -2 V can widen the hysteresis window, so the ON/OFF ratio can be enlarged. With $V_m$ increasing from 1 V to 2 V, the ON/OFF ratio range using the DC sweep from 4 V to -2 V is from $3.2 \times 10^3$ to $10^5$, and the maximum ON/OFF ratio value of $10^5$ is at $V_m = 1.8$ V. Meanwhile, the ON/OFF ratio using the DC sweep from 4 V to 0 V keeps decreasing from 914 to 12.6. It is seen that the DC sweep from 4 V to -2 V embraces a much higher ON/OFF ratio than the DC sweep from 4 V to 0 V.

Supplementary Fig. 4**b** shows the $V_m$ impact on the leakage power for a 32×32 (1 Kb) array. With $V_m$ increasing from 1 V to 2 V, the leakage power increases from 5 pW to 20 pW. In this article, the $V_m$ is set to 1.5 V with the consideration of the reliability, the ON/OFF ratio and the leakage power.

**The load line analysis on the a-IGZO MFMIS AFeFETs**

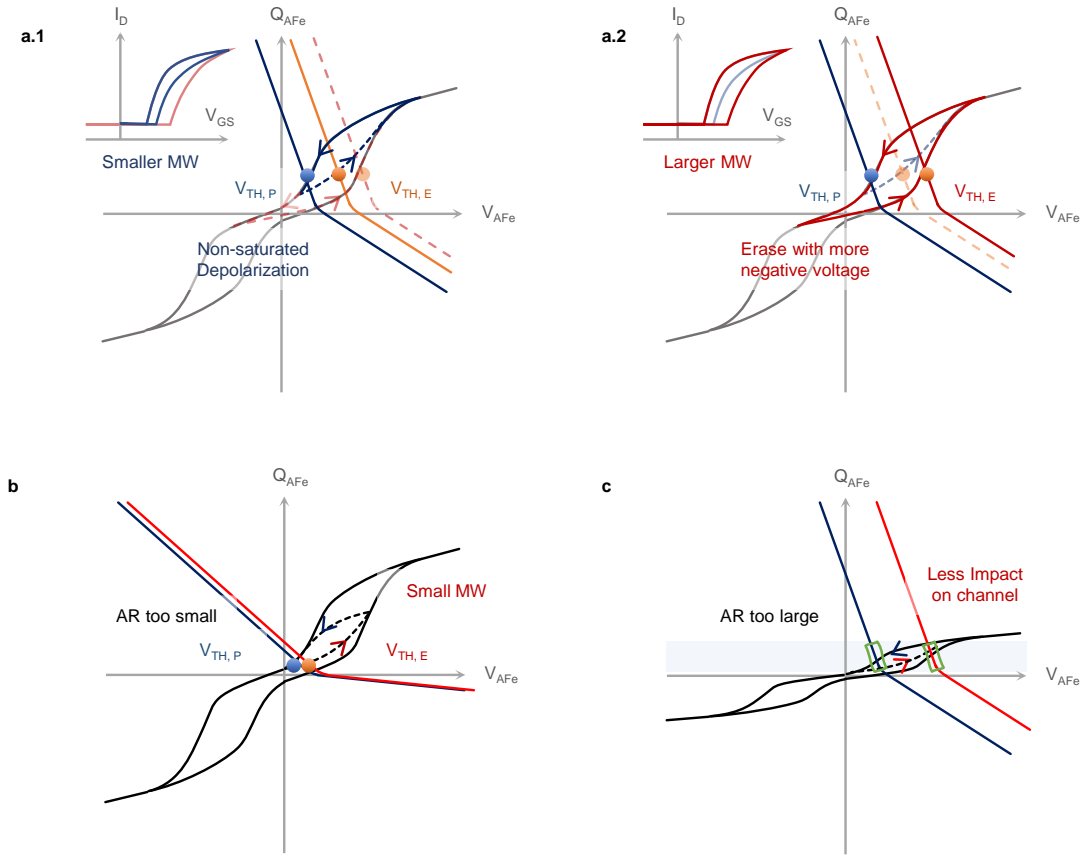

**Supplementary Fig. 7 | Load line analysis of the a-IGZO MFMIS AFeFETs. a**, Illustration of the MW enhancement with a negative erase voltage. Due to the non-zero flat-band voltage in the gate stack and inevitable charge trapping, it is possible that the AFE layer cannot be fully depolarized with unipolar operation, as shown in **a.1**. Here, typical load-line analysis with consideration of charge conservation is implemented with corresponding $I_D$-$V_{GS}$ curves in the subset. As observed in **a.2**, applying a negative gate bias can be required to depolarize the AFE layer; thus, the MW will increase from the blue curve in **a.1** to the red curve in **a.2**. **b**, The load line analysis for the AFeFETs with a smaller AR. If we keep the *P-V* of the capacitor constant, with a smaller AR, the *Q-V* of the MOS part will change, as shown in the figure. Thus, due to the less saturated switching of the AFE layer with the same operation voltage, the MW would consequently be smaller. **c**, The load-line analysis for the AFeFET with a larger AR. As mentioned in the main text for Fig. 2**d**, the devices with even larger ARs become hard to switch off within the sweeping range from 4 V to 0 V, and the ON/OFF ratios are also small. As depicted here, if we keep the *Q-V* of the MOS part constant, the *P-V* of the AFE layer would shrink accordingly. Furthermore, the region for the

operation point movement will be smaller (within the green boxes), meaning the carrier concentration in the channel will not change significantly with the given gate voltage.

## Variation issue and device scalability

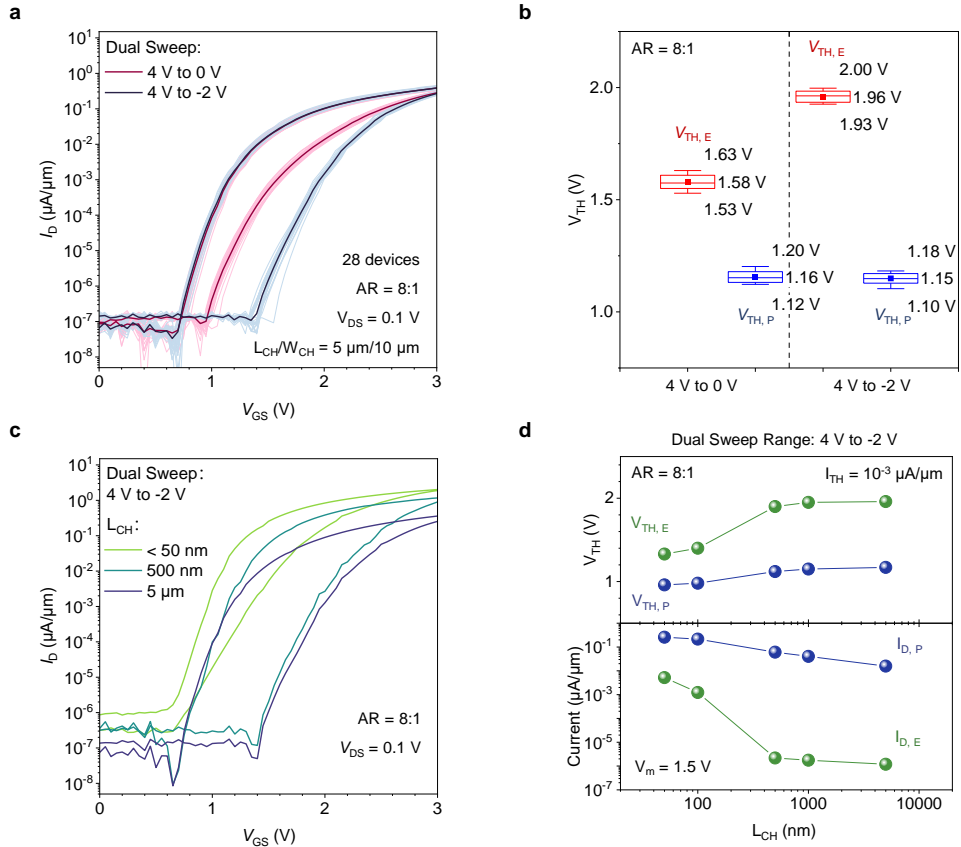

**Supplementary Fig. 8 | The variation of the a-IGZO AFeFETs and characteristics of scaled devices. a,** $I_D$-$V_{GS}$ curves of the 28 measured AFeFETs devices with AR of 8:1. **b,** Corresponding $V_{TH, E}$ and $V_{TH, P}$ box figures extracted from the $I_D$-$V_{GS}$ curves where the dots in the figure are the mean values and the boxes display the quartiles of the data. The devices are fabricated separately for variation validation, which can have a certain $V_{TH}$ shift compared with the other devices due to the process variation. For the devices with a smaller AR of 8:1, good device uniformity can be achieved with a standard deviation from 2.1% to 2.9% for each extracted $V_{TH}$. **c** and **d** are the $I_D$-$V_{GS}$ curves of the devices with various $L_{CH}$ and extracted $V_{TH}$ and $I_D$ for both program and erase states, respectively. The extraction methods are the same as those mentioned in the main text. Even with $L_{CH}$ down to less than 50 nm, the device still shows a proper function with anti-clockwise hysteresis within the $I_D$-$V_{GS}$ loops, which are induced by the AFE switching. The negatively-shifted $V_{TH}$ and MW degradation can be observed because the channel region becomes more difficult to deplete and the AFE switching needs more negative bias simultaneously. These can be improved with synergy thickness scaling for channel and dielectric layers within the gate stack.

## Endurance measurement

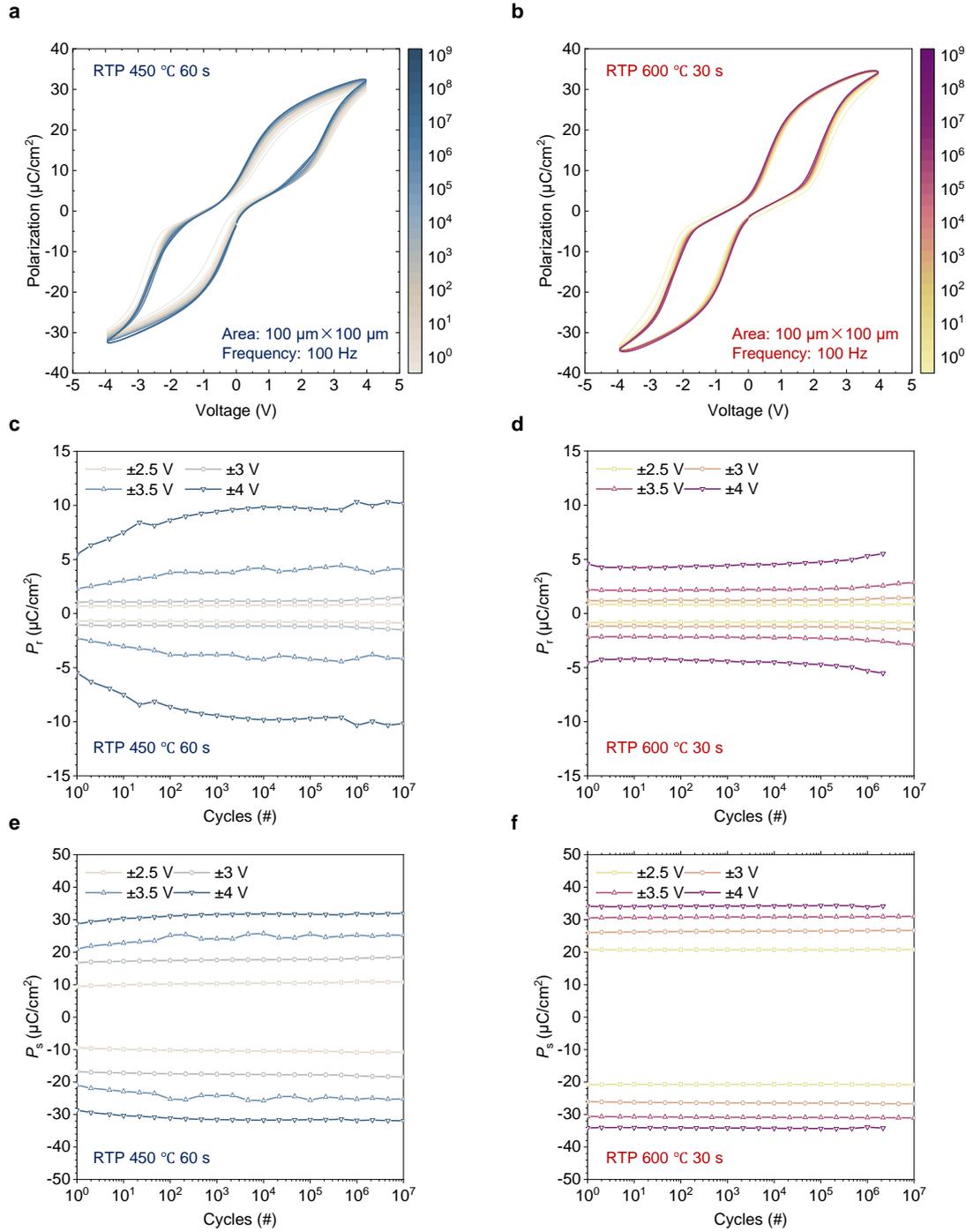

**Supplementary Fig. 9 | Endurance measurement of AFE capacitors with a MIM structure.**
**a**, **c** and **e** with curves in blue correspond to the capacitors with an RTP of 450 °C 60 s, while **b**, **d** and **f** with blue curves correspond to the capacitors with an RTP of 600 °C 30 s.

The devices are fabricated with the same process as the MIM part within the AFeFETs with an area of 100 μm×100 μm. **a** and **b** show the change of the $P$-$V$ curves with the cycling pulses of ±4 V. **c** and **d** show the extracted $P_r$ with different cycling voltages. **e** and **f** show the extracted $P_s$ with different cycling voltages. It can be observed from the figures that the AFE capacitors show excellent stability with negligible drifting for the *P-V* loops and nearly unchanged $P_r$ and $P_s$ during the endurance measurement. The results also indicate the BEOL compatibility of the $Hf_xZr_{1-x}O_2$-based AFE layer realizing strong AFE characteristics with an annealing temperature of only 450 °C. A higher annealing temperature and longer annealing time could result in more stable *P-V* characteristics of the capacitors, with eliminated stress-induced phase change from AFE to FE (gradually increased $P_r$ in **c**).

**Leakage overhead comparison**

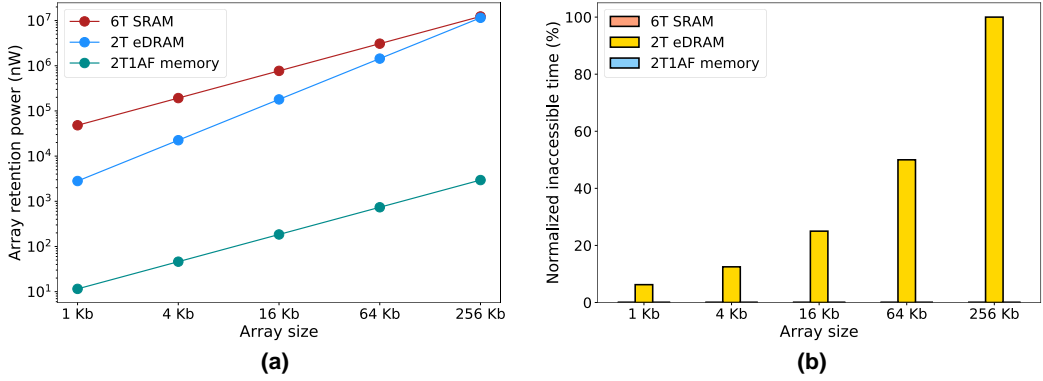

**Supplementary Fig. 10 | Leakage overhead comparison between the 6T SRAM, the 2T eDRAM, and the proposed 2T1AF memory. a**, The array retention power with different array sizes. **b**, The inaccessible time with different array sizes.

As illustrated in Introduction, volatile memories, including the CMOS SRAM and dynamic memories, suffer from the leakage problem, while the proposed 2T1AF memory shows outperformance with almost zero leakage. Leakage has a big impact on power consumption, and may degrade the memory performance when considering the leakage-induced refresh operations in dynamic memories. This figure compares the proposed 2T1AF memory with the conventional 6T SRAM and the 2T eDRAM on the retention power and inaccessible time with different array sizes. The array retention power considers the leakage power and the refresh power. The inaccessible time is the time that write and read requests must be stalled. Without loss of generality, the arrays are evaluated with the same number of rows and columns. The design parameters of the 6T SRAM and the 2T eDRAM are adopted from prior measurement results [31].

Supplementary Fig. 10**a** shows the array retention power comparison in variable array sizes. With the array size increasing from 1 Kb (64×64) to 256 Kb (512×512), the array retention power of the 6T SRAM and the 2T eDRAM increase from 48.1μW and 2.8μW to 12.3mW and 11.5mW, respectively. Note that the array retention power of the 2T eDRAM grows faster than that of the 6T SRAM. This is because the number of rows to be refreshed increases and the parasitic effect becomes more significant with a larger array size, while the retention power of the 6T SRAM array scales linearly with the array size. Meanwhile, the retention power of the 2T1AF memory array is only from 11.5 nW to 11.8 μW. The power savings over the 6T SRAM and the 2T eDRAM are up

to 4,178x and 7,805x, respectively. Such significant improvement is mainly from the eliminated subthreshold leakage in the LFVM operation.

Supplementary Fig. 10**b** illustrates the inaccessible time comparison of different memories in variable array sizes. It is seen that the inaccessible time of the 2T eDRAM increases linearly with the array size due to the increased row number, while the inaccessible time of the 6T SRAM and the 2T1AF memory is zero. By eliminating the refresh operation in the proposed LFVM, the system performance of the average memory access latency and instruction per cycle (IPC) can be improved significantly[22].